\documentstyle[twoside,titlepage,psfig,12pt]{article}
\begin{document}
\newcommand{\gtsima}{$\; \buildrel > \over \sim \;$}
\newcommand{\ltsima}{$\; \buildrel < \over \sim \;$}
\newcommand{\simgt}{\lower.5ex\hbox{\gtsima}}
\newcommand{\simlt}{\lower.5ex\hbox{\ltsima}}
\newcommand{\hikpc}{{\hbox {$h^{-1}$}{\rm kpc}} }
\newcommand{\himpc}{{\hbox {$h^{-1}$}{\rm Mpc}} }
\newcommand{\0}{ {\scriptscriptstyle {0}} }
\newcommand{\T}{ {\scriptscriptstyle {\rm T}} }
\newcommand{\kms}{ {\rm km/sec} }
\newcommand{\keV}{ {\rm keV} }
\newcommand{\mpc}{ {\rm Mpc} }
\newcommand{\bfs}{{\mbox{\boldmath $s$}}}
\newcommand{\bfx}{{\mbox{\boldmath $x$}}}
\newcommand{\bfk}{{\mbox{\boldmath $k$}}}
\newcommand{\smbfk}{{\mbox{\footnotesize\boldmath $k$}}}
\newcommand{\smbfx}{{\mbox{\footnotesize\boldmath $x$}}}
\newcommand{\sVert}{{\scriptscriptstyle\Vert}}
\def\pp{\par\parshape 2 0truecm 15.5truecm 1truecm 14.5truecm\noindent}

\begin{titlepage}
\vspace*{-1.5cm}
\begin{minipage}[c]{3cm}
  \psfig{figure=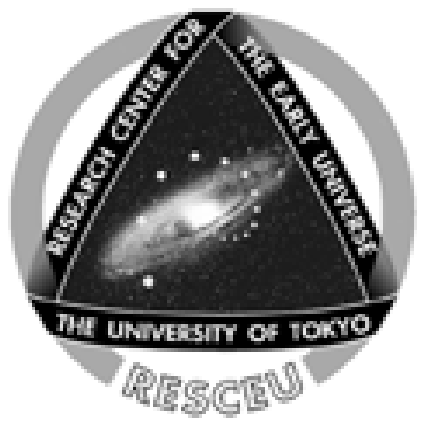,height=3cm}
\end{minipage}
\begin{minipage}[c]{10cm}
\begin{centering}
{
\vskip 0.1in
{\large \sf 
THE UNIVERSITY OF TOKYO\\
\vskip 0.1in
Research Center for the Early Universe}\\
}
\end{centering}
\end{minipage}
\begin{minipage}[c]{3cm}
\vspace{2.5cm}
RESCEU-11/96\\
UTAP-230/96
\end{minipage}\\
\vspace{1cm}

\addtocounter{footnote}{1}
 
\baselineskip=20pt

\begin{center}

  {\Large\bf Cosmological Redshift Distortion of Correlation
    Functions\\ as a Probe of the Density Parameter \\ and the
    Cosmological Constant}

\vspace{1cm}
 
{\sc Takahiko Matsubara and Yasushi Suto}

\end{center}
 

\noindent {\it Department of Physics, School
  of Science, The University of Tokyo, Tokyo 113, Japan.}

\noindent {\it Research Center For the Early Universe (RESCEU), 
  School of Science, The University of Tokyo, Tokyo 113, Japan.}


\centerline { e-mail: matsu@phys.s.u-tokyo.ac.jp, ~
suto@phys.s.u-tokyo.ac.jp}

\vspace{7mm}

\centerline{ApJL, in press.}

\vspace{1.0cm}

\baselineskip=14pt

\centerline {\bf Abstract}

\noindent
We propose cosmological redshift-space distortion of correlation
functions of galaxies and quasars as a probe of both the
density parameter $\Omega_0$ and the cosmological constant
$\lambda_0$. In particular, we show that redshift-space distortion of
quasar correlation functions at $z\sim2$ can in principle set 
a constraint on the value of $\lambda_0$. 
This is in contrast This is in
contrast to the popular analysis of galaxy correlation functions in
redshift space which basically determines $\Omega_0^{0.6}/b$, where
$b$ is the bias parameter, but is insensitive to $\lambda_0$. For
specific applications, we present redshift-space distortion of
correlation functions both in cold dark matter models and in power-law
correlation function models, and discuss the extent to which one can
discriminate between the different $\lambda_0$ models.

\medskip

\noindent {\it Subject headings:} cosmology: theory 
--- large-scale structure of the universe --- methods: statistical
\vfill

\end{titlepage}

\section{\normalsize\bf Introduction}

Redshift-space distortion of galaxy two-point correlation functions is
known as a powerful tool in estimating the cosmological density
parameter $\Omega_0$; on nonlinear scales Davis \& Peebles (1983)
computed the relative peculiar velocity dispersions of pair of
galaxies around $1h^{-1}$Mpc from the anisotropies in the correlation
functions of the CfA1 galaxy redshift survey, and then concluded that
$\Omega_0=0.20 e^{\pm0.4}$ (also see Mo, Jing \& B\"{o}rner 1993; Suto
1993; Ratcliffe et al. 1996). In linear theory Kaiser (1987) showed
that the peculiar velocity field systematically distorts the
correlation function observed in redshift space; the averaged
redshift-space correlation function $\xi^{(s)}(x)$ of galaxies is
related to its real-space counter part $\xi^{(r)}(x)$ as
\begin{eqnarray}
\label{eq:kaiser}
\xi^{(s)}(x) &=& \left( 1 + {2\over 3}\beta + {1\over 5}\beta^2 \right) 
\xi^{(r)}(x) , \\
\label{eq:beta}
\beta &=& { f(z=0) \over b}
   \sim {1\over b} 
\left[ \Omega_0^{0.6} + {\lambda_0 \over 70} 
\left(1+ {\Omega_0 \over 2}\right)\right] ,
\end{eqnarray}
where $b$ is the bias parameter, $f(z=0)$ is the logarithmic
derivative of the linear growth rate with respect to the scale factor
$a$ (eq.[\ref{eq:fz}] below) at $z=0$.  As is clear from the empirical
fitting formula in the second line by Lahav et al. (1991), however,
this formula depends on $\Omega_0$ but is practically insensitive to
the cosmological constant $\lambda_0$.

At higher redshift $z \simgt 1$, another anisotropy due to the
geometrical effect of the spatial curvature becomes important. We
derive a formula for {\it the cosmological redshift-space distortion}
in linear theory. This formula turns out to be a straightforward
generalization of that derived by Hamilton (1992) for $z=0$. It
provides a promising method to estimate both $\Omega_0$ {\it and}
$\lambda_0$ from future redshift surveys of galaxies and quasars
including the Sloan Digital Sky Survey (SDSS) and the 2dF. A very
similar idea was put forward independently by Ballinger, Peacock and
Heavens (1996) in which they considered the anisotropy of the power
spectrum while we developed a formulation in terms of two-point
correlation functions.  We outline the derivation in \S 2, and present
some examples of results in cold dark matter (CDM) models and in
power-law correlation function models.

\section{\normalsize\bf Cosmological redshift distortion}

Throughout the present analysis, we assume a standard Robertson --
Walker metric of the form:
\begin{equation}
ds^2 = -dt^2 + a(t)^2 
\{ d\chi^2 + S(\chi)^2 [d\theta^2 + \sin^2\theta d\phi^2 ] \} ,
\end{equation}
where $S(\chi)$ is determined by the sign of the spatial curvature $K$
as
\begin{equation}
 S(\chi) = 
  \left\{ 
      \begin{array}{ll}
         \sin{(\sqrt{K}\chi)}/\sqrt{K} & \mbox{$(K>0)$} \\
         \chi & \mbox{$(K=0)$} \\
         \sinh{(\sqrt{-K}\chi)}/\sqrt{-K} & \mbox{$(K<0)$} 
      \end{array}
   \right. 
 \\ \nonumber
\end{equation}
In our notation, $K$ is written in terms of the scale factor at
present $a_0$ and the Hubble constant $H_0$ as
\begin{equation}
K = a_0^2 H_0^2 (\Omega_0 + \lambda_0 -1) .
\end{equation}
The radial distance $\chi(z)$ is computed by
\begin{equation}
\chi(z) = \int_t^{t_0} {dt \over a(t)}  
 = { 1 \over a_0} \int_0^z 
{d z \over H(z)},
\end{equation}
where we introduce the Hubble parameter at redshift $z$:
\begin{equation}
   H(z) = H_0\sqrt{\Omega_0 (1 + z)^3 +
(1-\Omega_0-\lambda_0) (1 + z)^2 + \lambda_0} .
\end{equation}

Consider a pair of galaxies or quasars located at redshifts $z_1$ and
$z_2$.  If both the redshift difference $\delta z \equiv z_1-z_2$ and
the angular separation of the pair $\delta\theta$ are much less than
unity, the comoving separations of the pair parallel and perpendicular
to the line-of-sight direction, $x_{\sVert}$ and $x_\bot$, are given
by
\begin{equation}
\label{eq:xdef2}
x_{\sVert} (z) = {d\chi(z)\over dz}  \delta z =
 \frac{c_{\sVert} \delta z}{a_0H_0}, \quad
x_\bot (z)  = S(\chi(z)) \delta \theta =
\frac{c_\bot z \delta \theta}{a_0H_0},
\end{equation}
where $c_\sVert (z) = H_0/H(z)$, $c_\bot (z) = a_0 H_0 S(\chi(z))/z$
and $z \equiv (z_1+z_2)/2$. Thus their ratio becomes
\begin{eqnarray} 
\label{eq:ratio}
{x_\sVert (z) \over x_\bot (z) } 
= \frac{c_\sVert(z)}{c_\bot (z)}{\delta z \over z \delta \theta}
\equiv \eta(z) {\delta z \over z \delta \theta} .
\end{eqnarray}
Since $x_\sVert/x_\bot$ should approach $\delta z/(z \delta \theta)$
for $z \ll 1$, $\eta(z)$ can be regarded to represent the correction
factor for the {\it cosmological redshift distortion} (upper panel of
Fig.1). This was first pointed out in an explicit form by Alcock \&
Paczy\'nski (1979).

Although $\eta(z)$ is a potentially sensitive probe of $\Omega_0$ and
especially $\lambda_0$, this is not directly observable unless one has
an independent estimate of the ratio $x_\sVert/x_{\bot}$.  Therefore
Alcock \& Paczy\'nski (1979) proposed to apply the result to
intrinsically spherical structure resulting from gravitational
clustering. More recently Ryden (1995) proposed to apply the test to
statistically spherical voids around $z\simlt 0.5$. It is not likely,
however, that the intrinsic shape of the objects formed via
gravitational clustering is spherical even in a statistical
sense. Thus their proposal would be seriously contaminated by the
aspherical shape distribution of the objects considered.  In what
follows, we propose to use the clustering of quasars and galaxies in
$z$, which should be safely assumed to be statistically isotropic, in
estimating $\eta(z)$. Also we take account of the distortion due to
the peculiar velocity field using linear theory.

If an object located at redshift $z_1$ has a peculiar recession
velocity $v_{\sVert,1}$ with respect to us, we actually {\it observe}
its redshift as
\begin{equation}
   1 + z_{\rm obs,1} = (1 + z_1)(1 + v_{\sVert,1} - v_{\sVert,0})
\end{equation}
where $v_{\sVert,0}$ is our (observer's) peculiar velocity, and we
assume that all peculiar velocities are non-relativistic.  Let us
consider a sample of galaxies or quasars located in the range $z-dz
\sim z+dz$ with $dz\ll z$ distribute over an angular radius comparable
to the proper distance corresponding to $dz$. Then we set the locally
Euclidean coordinates in real space (comoving), $\bfx$, and in
redshift space, $\bfs$, which share the same origin at redshift $z$.
Since we assume that these coordinates are far from us, we can use the
distant-observer approximation. If we take the third axis along the
line of sight, the above two coordinates are related to each other as
\begin{eqnarray}
   &&
   s_1 = \frac{x_1}{c_\bot (z)},\quad  s_2 = \frac{x_2}{c_\bot (z)},
   \\
   &&
   s_3 = \frac{z_{\rm obs,1} - z}{H_0} \simeq
   {1 \over c_\sVert(z)} \left[x_3 + 
          \frac{1 + z}{H(z)}(v_{\sVert} - v_{0\sVert}) \right].
\end{eqnarray}
Note that in the last expression, we assume that $x_3$ is sufficiently
small and replace $1 + z + H x_3$ by $1 + z$. The number densities of
objects in these coordinates are related by the Jacobian of the above
transformation. In linear order of density contrast and peculiar
velocity, we obtain
\begin{equation}
   \delta^{(s)} (\bfs(\bfx)) = 
   \delta^{(r)} (\bfx) -  \frac{1 + z}{H(z)} \partial_3 v_\sVert.
   \label{eqm1}
\end{equation}
According to linear theory, the peculiar velocity is related to the
{\it mass} density fluctuation $\delta_{\rm m}$ (e.g., Peebles 1993)
as
\begin{equation}
  v_\sVert (\bfx) = - \frac{H(z)}{1 + z} f(z) \partial_3 \triangle^{-1}
  \delta_{\rm m}(\bfx),
\end{equation}
where $\triangle^{-1}$ is the inverse Laplacian,
\begin{eqnarray} 
\label{eq:fz}
   f(z) \equiv \frac{d\ln D(z)}{d\ln a} \simeq
   \Omega(z)^{0.6} + {\lambda(z) \over 70} 
   \left(1+ {\Omega(z) \over 2}\right),
\end{eqnarray}
and the linear growth rate $D(z)$ (normalized as $D(z)=1/(1+z)$ for
$z\rightarrow \infty$) is
\begin{equation}
   D(z) = {5\Omega_0H_0^2 \over2}H(z)
   \int_z^\infty {1+z' \over H(z')^3}\, dz'.
\end{equation}
In the above, $\Omega(z)$ and $\lambda(z)$ are the density parameter
and the dimensionless cosmological constant at $z$:
\begin{equation} 
   \Omega(z) = \left[{H_0 \over H(z)}\right]^2 \, (1+z)^3 \Omega_0 ,\quad
   \lambda(z) = \left[{H_0 \over H(z)}\right]^2 \, \lambda_0.
\end{equation}
Allowing for the possibility that the density contrast of {\it mass},
$\delta_{\rm m}$, differs from that of {\it objects} $\delta^{(r)}$,
we assume linear biasing $\delta^{(r)} = b\: \delta_{\rm m}$, where
$b(z)$ is a bias factor which now depends on $z$. Then equation
(\ref{eqm1}) simply generalizes Kaiser (1987)'s result to $z \neq 0$:
\begin{equation}
\label{eq:kaiserz}
   \delta^{(s)}(\bfs(\bfx)) =
   \int \frac{d^3k}{(2\pi)^3}
   \left[1 + \beta(z) \frac{k_3^{\;2}}{k^2} \right]
   e^{i \smbfk \cdot \smbfx}
   {\widetilde \delta}^{(r)}(\bfk),
\end{equation}
where $\beta(z) = f(z)/b(z)$, and ${\widetilde \delta}^{(r)}$ is the
Fourier transform of the density contrast. From equation
(\ref{eq:kaiserz}), we derive the following relation for the two-point
correlation function at $z \neq 0$ which also generalizes the formula
of Hamilton (1992) at $z=0$:
\begin{eqnarray} 
\label{eq:xis}
   &&
   \xi^{(s)}(s_\bot, s_\sVert) = 
   \left(1+{2\over 3}\beta(z) +{1\over5}[\beta(z)]^2\right) \xi_0(x)
   P_0(\mu) 
   -\left({4\over 3}\beta(z)
   +{4\over7}[\beta(z)]^2\right)\xi_2(x) P_2(\mu)
   \nonumber\\
   && \qquad\qquad
   +\frac{8}{35}[\beta(z)]^2 \xi_4(x) P_4(\mu) ,
\end{eqnarray}
where $x \equiv \sqrt{{c_\sVert}^2 {s_\sVert}^2 + {c_\bot}^2
  {s_\bot}^2}$, $\mu\equiv c_{\sVert} s_{\sVert} /x$ ($s_\sVert=s_3$
and $s_\bot^2 = s_1^2+s_2^2$), $P_n$'s are the Legendre polynomials,
and
\begin{eqnarray} 
\label{eq:xi2l}
\xi_{2l}(x) &\equiv& {1 \over 2\pi^2} 
   \int_0^\infty dk k^2 j_{2l}(kx) P(k;z)  \cr
&=& { (-1)^l \over x^{2l+1}} \left(\int_0^x xdx\right)^l x^{2l}
\left({d \over dx}{1 \over x}\right)^l x \xi^{(r)}(x;z) .
\end{eqnarray}
Again in linear theory and for linear biasing, the power spectrum
$P(k;z)$ of objects at $z$ is related to that of {\it mass}
$P^{(m)}(k)$ of matter at $z=0$ as
\begin{equation}
   P(k;z) = [b(z)]^2\left(\frac{D(z)}{D(0)}\right)^2
   P^{(m)}(k),
\end{equation}

\section{\normalsize\bf Simple model predictions}

The lower panel in Figure 1 plots $f(z)$. Here and in what follows, we
use the fitting formula (\ref{eq:fz}) for $f(z)$ whose mean error is 2
percent.  As expected from equation (\ref{eq:beta}), $f(z=0)$ is
insensitive to the value of $\lambda_0$ and basically determined by
$\Omega_0$ only. At higher redshifts, however, $f(z)$ becomes
sensitive to $\lambda_0$ as pointed out earlier by Lahav et al (1991),
especially if $\Omega_0 \ll 1$.  Therefore the behavior of $f(z)$ at
low and high $z$ is a potentially good discriminator of $\Omega_0$ and
$\lambda_0$, respectively. What we propose here is that a careful
analysis of the redshift distortion in correlation functions of
galaxies at low redshifts and quasars at high redshifts can probe
$\Omega_0$ and $\lambda_0$ through $f(z)$ as well as $\eta(z)$.

For specific examples, we compute $\xi^{(s)}(s_\bot,
s_{\sVert})$ in linear theory applying equations
(\ref{eq:xis}) and (\ref{eq:xi2l}) in CDM models with $H_0=70$
km$\cdot$s$^{-1}\cdot$Mpc$^{-1}$ (Tanvir et al. 1995). The resulting
contours are plotted in Figure 2. The four sets of values of
$\Omega_0$ and $\lambda_0$ are the same as adopted in Figure 1, and we
adopt the COBE normalization (Sugiyama 1995; White and Scott 1995).

In order to quantify the cosmological redshift distortion in Figure 2,
let us introduce the anisotropy parameter
$\xi^{(s)}_{\sVert}(s)/\xi^{(s)}_\bot(s)$, where $\xi^{(s)}_\bot(s)
\equiv \xi^{(s)}(s,0)$ and $\xi^{(s)}_{\sVert}(s) \equiv
\xi^{(s)}(0,s)$. The left four panels in Figure 3 show the anisotropy
parameter against $z$ in CDM models. Since the evolution of bias is
largely unknown, we assume $b = 1$ and $2$ just for definiteness. The
effect of the evolution of bias (e.g., Fry 1996) will be considered
elsewhere (Suto \& Matsubara 1996). This clearly exhibits the extent
to which one can discriminate the different $\lambda_0$ models on the
basis of the anisotropies in $\xi^{(s)}$ at high redshifts.

For comparison, let us consider a simple power-law model
$\xi^{(r)}(x;z) = A(z) x^{-\gamma}$ ($\gamma>0$). Then equation
(\ref{eq:xis}) reduces to
\begin{eqnarray} 
\label{eq:xisplaw}
{\xi^{(s)}(s_\bot, s_{\sVert}) \over
\xi^{(r)}(\sqrt{c_{\sVert}^2
    s_{\sVert}^2 + c_\bot^2 s_\bot^2};z) }
= 1 + { 2(1-\gamma\mu^2) \over 3-\gamma} \beta(z) 
+ { \gamma(\gamma+2)\mu^4-6\gamma\mu^2+3
\over (3-\gamma)(5-\gamma)} \beta(z)^2 .
\end{eqnarray}
In this case the anisotropy parameter 
is given by
\begin{eqnarray} 
{\xi^{(s)}_{\sVert}(s) \over \xi^{(s)}_\bot(s)}
= \eta(z)^{-\gamma} 
{ (3-\gamma)(5-\gamma) + 2(1-\gamma)(5-\gamma)\beta(z) 
+ (3-\gamma)(1-\gamma) \beta(z)^2 
\over
(3-\gamma)(5-\gamma) + 2(5-\gamma)\beta(z) + 3\beta(z)^2  } ,
\end{eqnarray}
and is independent of the scale $s$.  

The right four panels in Figure 3 show that the behavior of the
anisotropy parameter is sensitive to the power-law index
$\gamma$. This is partly because $\Omega_0=1$ CDM models which we
consider are already nonlinear on $(10\sim20)\himpc$ (Fig.2), and then
the linear theory prediction is not reliable enough.  On the other
hand, $\Omega=0.1$ CDM models are well described in linear theory on
the scales of our interest $\simgt 1\himpc$.  Since most observations
do suggest that $\Omega_0$ in our universe is around $(0.1\sim 0.3)$
(e.g., Peebles 1993; Suto 1993; Ratcliffe et al. 1996), this is
encouraging for our present result on the basis of linear theory.
Incidentally, Figure 3 also implies that $\Omega=0.1$ CDM models in
the linear regime are well approximated by the power-law model with
$\gamma \sim 1$. This is reasonable since the CDM spectral index of
$P(k)$ is around $-2$ which corresponds to $\gamma\sim1$.

\section{\normalsize\bf Conclusions}

Redshift-space distortion of galaxy correlation functions has
attracted much attention as a tool to determine $\Omega_0^{0.6}/b$ (Kaiser
1987; Hamilton 1992). We obtained an expression to describe the
cosmological redshift-space distortion at high $z$ in linear theory
taking proper account of the spatial curvature $K$.  Then we showed
that in principle this can discriminate the value of $\lambda_0$ via the
$z$-dependence of the $\beta(z)$ and $\eta(z)$ parameters.

Recent analysis, for instance of the Durham/UKST galaxy redshift
survey on the basis of eq.(\ref{eq:kaiser}), yields $\Omega_0^{0.6}/b
= 0.55 \pm 0.12$ (Ratcliffe et al. 1996) with $\sim 2500$
galaxies. Their Figure 4(a) clearly exhibits that the observational
data at $z=0$ are already statistically reliable for the direct
quantitative comparison with the upper panels in our Figure 2,
although we do not attempt it at this point. 
SDSS, for instance, is
expected to have a million of galaxy redshifts up to $z\simlt 0.2$
which would be able to determine $\Omega_0^{0.6}/b$ to better than 10
percent accuracy.  Again upon completion of SDSS, $\sim 10^5$ quasar
samples become available and the anisotropy in quasar correlation
functions at $z\sim2$ will put a constraint on $\lambda_0$, given
$\Omega_0$ determined from the galaxy correlation functions. 
The quasar luminosity function of Boyle, Shanks \& Peterson (1988)
predicts that the number of quasars per $\pi$ steradian brighter than
19 B magnitude is about 4500 either in $z=0.9\sim 1.1$ or in
$z=1.9\sim 2.1$ (for the $\Omega_0=1$ and $\lambda_0=0$
model). Therefore we expect that the resulting statistics is even
better than that obtained by Ratcliffe et al. (1996) if unknown
evolution effects and other potential systematics interfere.

Although the linear theory which we used throughout the paper becomes
less restrictive at higher redshifts, it is observationally easier to
detect clustering features in the nonlinear regime. Thus the analysis
of nonlinear redshift-space distortion is another important area for
research (Suto \& Suginohara 1991; Matsubara \& Suto 1994; Suto \&
Matsubara 1994; Matsubara 1994).  In addition, it is important to
examine the possible systematic errors due to the finite volume size
and the shape of the survey region. This can be best investigated by
the direct analysis of the numerical simulations. This is partly
considered by Ballinger, Peacock, \& Heavens (1996) although in
k-space. We plan to return to this in the later paper (Magira,
Matsubara, \& Suto 1996).

\bigskip

We thank Naoshi Sugiyama for discussions.  This research was supported
in part by the Grants-in-Aid by the Ministry of Education, Science and
Culture of Japan (07740183, 07CE2002).  After we submitted the present
paper, we realized the similar independent work by Ballinger, Peacock,
\& Heavens (1996).  We are grateful to John Peacock for calling our
attention to this paper and for useful comments.

\bigskip
\bigskip
\newpage

\parskip2pt
\bigskip
\centerline{\bf REFERENCES}
\bigskip

\def\apjpap#1;#2;#3;#4; {\pp#1, {#2}, {#3}, #4}
\def\apjbook#1;#2;#3;#4; {\pp#1, {#2} (#3: #4)}
\def\apjppt#1;#2; {\pp#1, #2.}
\def\apjproc#1;#2;#3;#4;#5;#6; {\pp#1, {#2} #3, (#4: #5), #6}

\apjpap Alcock, C. \& Paczy\'nski, B. 1979;Nature;281;358;
\apjppt Ballinger, W.E., Peacock, J.A. \& Heavens, A.F. 1996;MNRAS, in
press;
\apjpap Boyle, B.J., Shanks, T. \& Peterson, B.A. 1988;MNRAS;235;935;
\apjpap Davis, M. \& Peebles, P.J.E. 1983;ApJ;267;465;
\apjpap Fry, J. 1996;ApJ;461;L65;
\apjpap Hamilton, A.J.S. 1992;ApJL;385;L5;
\apjpap Kaiser, N. 1987;MNRAS;227;1;
\apjppt Magira,~H., Matsubara,~T. \& Suto,~Y.;in preparation.;
\apjpap Matsubara,~T. 1994;ApJ;424;30;
\apjpap Matsubara,~T. \& Suto, Y.  1994;ApJ;420;497;
\apjpap Mo, H.J., Jing, Y.P., \& B\"{o}rner, G. 1993;MNRAS;264;825;
\apjpap Lahav, O., Lilje, P.B., Primack, J.R., \& Rees, M.J. 1991;
MNRAS;251;128;
\apjbook Peebles,P.J.E. 1993; Principles of Physical Cosmology;
    Princeton University Press;Princeton;
\apjppt Ratcliffe,A. et al. 1996;preprint, astro-ph/9602062;
\apjpap Ryden,B.1995;ApJ;452;25;
\apjpap Sugiyama N. 1995; ApJS; 100; 281; 
\apjpap Suto, Y.  1993;Prog.Theor.Phys.;90;1173;
\apjpap Suto, Y. \& Matsubara,~T.  1994;ApJ;420;504;
\apjpap Suto,~Y., \& Suginohara,~T. 1991;ApJ;370;L15;
\apjpap Tanvir, N.R., Shanks, T., Ferguson, H.C. \& Robinson,
D.R.T. 1995; Nature; 377; 27;
\apjpap White, M. \& Scott, D. 1996;ApJ;459;415;

\bigskip
\bigskip
\newpage

\bigskip 

\pp {\bf Figure 1 :} $\eta(z)=c_\sVert(z)/c_\bot (z)$ ({\it upper
  panel}), and $f(z)=\beta(z)b(z)$ ({\it lower panel}) for
($\Omega_0$, $\lambda_0$) = (1.0,0.0) in {\it dashed line}, (1.0,0.9)
in {\it dot-dashed line}, (0.1,0.9) in {\it dotted line}, and
(1.0,0.9) in {\it thick solid line}.

\bigskip

\pp {\bf Figure 2 :} The contours of $\xi^{(s)}(s_\bot, s_{\sVert})$
in CDM models at $z=0$ ({\it upper panels}) and $z=3$ ({\it lower
  panels}).  $H_0=70$km$\cdot$s$^{-1}\cdot$Mpc$^{-1}$ is assumed, and
the amplitude of the fluctuation spectrum is normalized according to
the COBE 2 yr data.  Solid and dashed lines indicate the positive and
negative $\xi^{(s)}$, respectively.  Contour spacings are $\Delta {\rm
  log}_{10} |\xi| = 0.25$. Thick lines in all $\Omega_0=1$ models
represent $\xi^{(s)}=10$, $1$, $0.1$, $-0.1$, and $-1$.  Thick lines
in $\Omega_0=0.1$ models indicate that $\xi^{(s)}=0.01$ and $0.001$
for $\lambda_0=0$ at $z=0$, $\xi^{(s)}=1$, $0.1$ and $0.01$ for
$\lambda_0=0.9$ at $z=0$, $\xi^{(s)}=0.01$ for $\lambda_0=0$ at $z=3$,
and $\xi^{(s)}=0.1$ for $\lambda_0=0.9$ at $z=3$.

\bigskip 

\pp {\bf Figure 3 :} The anisotropy parameter
$\xi^{(s)}_{\sVert}(s)/\xi^{(s)}_\bot(s)$ as a function of $z$.  Upper
and lower panels assume that $b=1$ and $2$, respectively.  {}From left
to right, the panel corresponds to CDM at $s=10\himpc$, CDM at
$s=20\himpc$, a power-law model with $\gamma=1.8$ and a power-law
model with $\gamma=1.0$.

\end{document}